\documentclass[aps,pre,twocolumn,superscriptaddress]{revtex4} 
\usepackage{graphicx,amsmath}
\usepackage[binary,amssymb]{SIunits}
\usepackage{color}
\usepackage{ulem}
\usepackage{placeins}
\usepackage{tikz}  
\newcommand{\makered}[1]{\textcolor{red}{#1}}
\renewcommand{\makered}[1]{#1}
\newcommand{\makeredred}[1]{\textcolor{red}{#1}}
\renewcommand{\makeredred}[1]{#1}

\DeclareRobustCommand\circled[1]{%
            \tikz[
              baseline=-1mm,
            ]{%
                \clip (0mm,-0.05mm) circle [radius=1.55mm];
                \draw[black, line width=0.2mm]
                   (0mm,-0.05mm) circle [radius=1.5mm];%
                \node[black, anchor=center, scale=0.75, yshift=0.05mm]
                (0mm,0mm) {#1};%
            }}%

\begin{document}

\title{Using ray-traversal for 3D particle matching\\in the context of particle tracking velocimetry \makeredred{in fluid mechanics}}
\author{Micka\"el Bourgoin}
\email{mickael.bourgoin@ens-lyon.fr}
\affiliation{Univ Lyon, \'Ecole normale sup\'erieure de Lyon, Univ Claude Bernard Lyon 1, C.N.R.S., Laboratoire de Physique, F-69342 Lyon, France}
\author{Sander G. Huisman}  
\email{s.g.huisman@gmail.com}
\affiliation{Physics of Fluids Group, Max Planck UT Center for Complex Fluid Dynamics, Faculty of Science and Technology, MESA+ Institute, and J.M. Burgers Centre for Fluid Dynamics, University of Twente, P.O. Box 217, 7500 AE Enschede, The Netherlands}
\affiliation{Univ Lyon, \'Ecole normale sup\'erieure de Lyon, Univ Claude Bernard Lyon 1, C.N.R.S., Laboratoire de Physique, F-69342 Lyon, France}

\date{\today}

\begin{abstract} 
An innovative method based on the traversal of rays, originating from detected particles, through a three-dimensional grid of voxels is presented. The methodology has as main advantage that the outcome of the method is independent of the order of the input; the order of the cameras and the order of the rays presented as input to the algorithm does not influence the outcome. The algorithm finds matches in decreasing value of match quality, ensuring that globally best matches are matched before \makered{worse} matches. The \makered{time complexity of the} algorithm is found to scale efficiently with the number of cameras and particles. \makered{A variety of show-cases are given to exemplify the algorithm for different geometries and different number of cameras.} \makeredred{The method is designed for the tracking of tracer or inertial particles in fluid mechanics, for which the particle size generally ranges from $\mathcal{O}(\micro\meter)$--$\mathcal{O}(\centi\meter)$. The method, however, does not impose a size limit on the particles.}
\end{abstract}

\maketitle

\section{Introduction}
The process of calculating the 3D position based on the views of multiple cameras is traditionally called stereomatching, and is based on the biological process of \textit{stereopsis}. Though most animals have binocular vision, experiments in fluid dynamics have been using more than two cameras to improve depth perception and to provide extra robustness and accuracy by data-redundancy, in particular in the context of Lagrangian particle tracking techniques \cite{maas,virant,ott,bourgoin2006} \makeredred{(with particles generally having a size in the range of $\micro\meter$--$\centi\meter$)}. The angles between the cameras should, however, be optimized. A small angle between two cameras causes large errors in the estimation of the depth. It is fairly common to put cameras at relative angles of 90 degrees, such that one of the coordinates is redundant, which makes matching easier. It is, though, not strictly necessary; it can be any angle. When more than two cameras are used, in order to minimize the error in all directions, it is generally a good strategy to globally maximise the relative angles between all the cameras. 

\makered{In the past years, important advancements have been done in the field of 3D multi-view reconstruction, impulsed by progresses in computer vision science \cite{hartley2003}. In particular, situations where cameras record projective images (with a linear correspondence between real world views and acquired images) have reached a high level of mathematical understanding and algorithm developments which allow to perform 3D-matching from multi-camera recordings, with a minimum requirement of \textit{a priori} knowledge of the camera arrangement and optical properties (so called internal and external calibration parameters)
\cite{zhang1995,hartley1995,criminisi2001}. In this regard the 3D reconstruction from multiple projective views is generally considered as a solved problem \cite{hartley2003}. The case of non-projective views has, however, advanced less. This is a common situation in fluid mechanics research applications, where images can be taken through multiple interfaces, with different shapes (non-necessarily planar), separating different media (with different optical properties, as air-water interfaces) \textit{et cetera}, what eventually lead to strong refraction effects and non-linear distorsions. For this reason, 3D-view reconstruction in high-resolution fluid mechanics measurements (for instance in \textit{Particle Tracking Velocimetry} (PTV)) still generally relies on accurate camera calibration methods capable to handle such non-linearities \cite{basanta2013,machicoane} and to retrieve accurate correspondences for each camera between each pixel and the corresponding ray of light that produces an image on the pixel. A 3D-matching algorithm then proceeds by seeking correct correspondences of rays between multiple cameras.}

\begin{figure}[h!]
	\begin{center}
		\includegraphics[width=0.7\columnwidth]{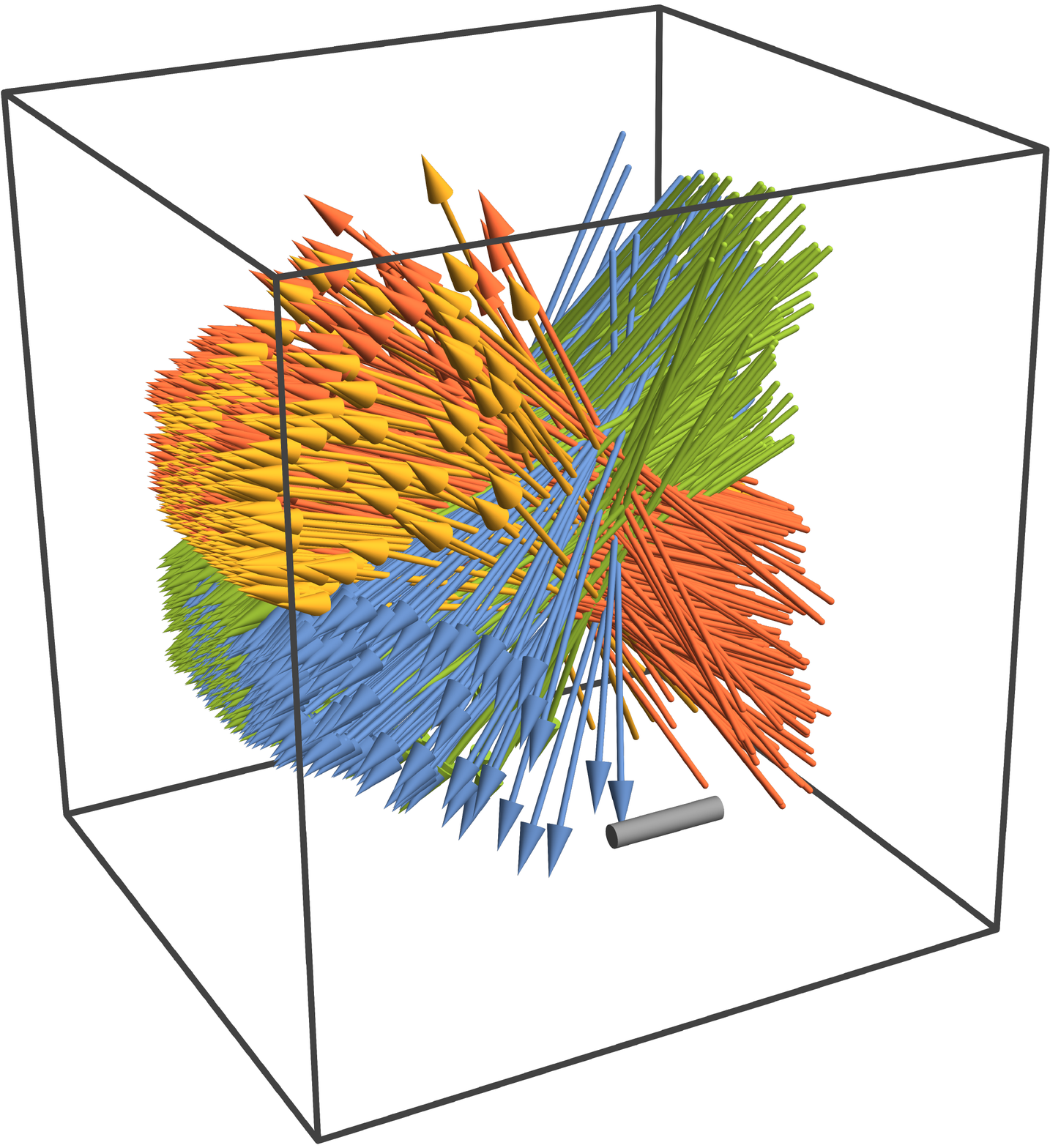}
		\caption{Example of rays for the setup shown in Fig.~\ref{fig:examples}a. Gray scale bar has a length of \unit{50}{\milli\meter}. A total of $\sim 1600$ rays intersect a measurement volume. Rays originating from the same camera have the same color. Data taken from Ref.~\cite{mathai2018}.}
		\label{fig:rayintersection}
	\end{center} 
\end{figure}

This manuscript is on the matching algorithm of light rays from a set of several cameras used for 3D particle tracking methods. In this context, the 3D position of a particle is retrieved from the intersection of rays of light coming from each of the cameras. These rays can be obtained e.g.~from Tsai's pinhole model\cite{tsai} based on the optics of the camera and the objectives or more elaborate models\cite{machicoane}. Like ray-tracing in computer graphics, we will consider the rays to origin from each camera, towards the detected particle---in the opposite direction of light being scattered by a particle that is captured by a camera. 
For a given particle and a given camera, this ray $r$ has an origin $p$ (let's say at the position of the particle image on the camera sensor) and a direction $v$. The complexity of the problem arises when many particles are to be tracked simultaneously with several cameras. In this situation, a bundle of rays (one ray for each particle) emerges from each camera. In order to determine the actual 3D position of all the particles, one needs to find the \makered{set of rays} that cross (or nearly cross) with each other. To get a feeling of the difficulty of the problem, we show an example set of rays from experiments having slightly over 1600 rays in total, emerging from 4 cameras at different view angles, see Fig.~\ref{fig:rayintersection}.
The problem of 3D matching can then be stated as follows: given $c$ cameras, each with $m_i$ rays, find sets of rays $r_{i.j}$ that minimize the distance from a point to  rays coming from different cameras. Here the $i$ index is the index of the camera and $j$ the index of the ray for that camera. 

For a valid match, the rays from each set of rays should be from distinct cameras. A simple \makered{na\"ive} approach would be to consider all possible sets and pick out the best matches\makered{, i.e.~a brute force computation.} However this would lead to $m^c$ combinations to be tested. For a typical case of $400$ particles, tracked with $4$ cameras, this would lead to a large number of candidate matches ($400^4 = 2.56 \times 10^{10}$). This is computationally prohibitive as too much time would be spent to go through all combinations looking for crossing or nearly crossing lines. Various strategies can be thought of that will eliminate the majority of these candidate matches. Classical alternative strategies are generally based on \makered{epipolar geometry} with projections of rays between pairs of cameras~\cite{maas,ott,bib:risoeReport}, in order to reduce the dimension of space in which crossings are to be found. The most efficient schemes can reduce the computational need from $\mathcal{O}(m^c)$ to $\mathcal{O}(cm\log{m})$. However, such strategies usually operate by successive stereo-matching searches of correspondences between pairs of cameras. When more than two camera are used, this requires then either to consider one of the cameras as a reference for the pairs (with time complexity $\mathcal{O}(cm\log{m})$), what may lead to ambiguities (as due to imperfection of the optical models, the matches found may depend on the choice of the reference camera) or to consider all possible pairs of cameras (with time complexity $\mathcal{O}(c^2m\log{m})$) and apply sophisticated combinatory algorithms to perform the required consistency checks between all the pairwise stereo-matches to avoid ambiguities. \makered{Multi-focal geometry offers an alternative robust and efficient framework to achieve this, by building a set of linear tensors directly connecting the views from multiple cameras \cite{hartley1995,hartley2003}. However this linear approach is strongly tight to the projective model generally used to describe the cameras behavior and can hardly be extend to more general situations where non-linear corrections (accounting for instance for optical distortions) are necessary.}

Here we propose a new strategy, based on ray traversal across 3D voxels, which efficiently allows to perform the stereo-matching with an arbitrary number of cameras, by combining all the cameras simultaneously, without requiring pair-wise operations and consistency post-checks \makered{and independently of the calibration model used to construct the rays, which can be as simple as a pinhole camera model \cite{tsai} or any more sophisticated non-linear calibration \cite{machicoane} giving the pixel-to-ray correspondence}. \makered{Note that we will focus here on applications of the newly proposed matching method to PTV in fluid mechanics; we will not discuss the calibration and tracking (\textit{i.e.}~the following of matched particles over time) parts of this technique, as they can be achieved independently}. Beyond PTV, the present method could also be used for the 3D reconstruction of an object by matching image keypoints from multiple images from different angles. 

\section{Method}
\begin{figure}[bht]
	\begin{center}
		\includegraphics{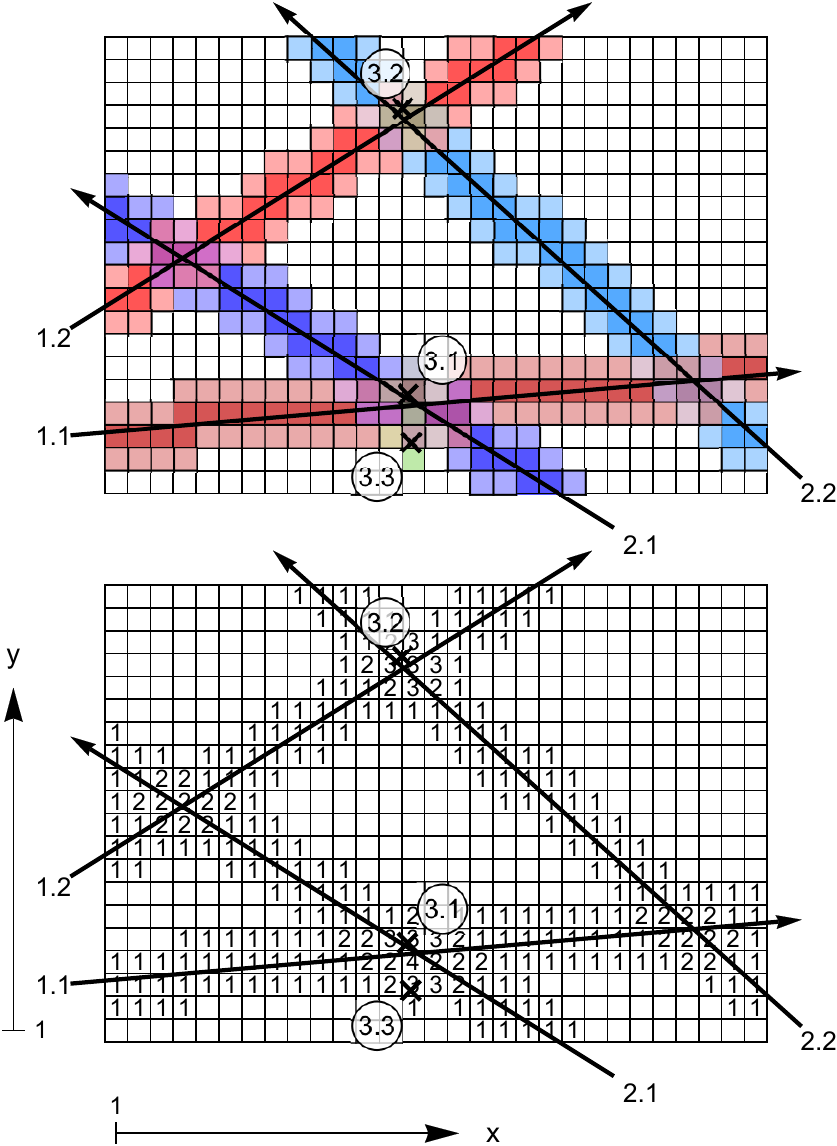}
		\caption{Top: 7 rays of light are traversing through a 2D slice of a 3D voxel-array, rays 1.1, 1.2, 2.1, and 2.2 are moving parallel to the plane of the slice, while 3.1, 3.2, and 3.3 (denoted by a $\times$) are perpendicular to plane. We expand the traversed voxels one neighbour in each direction, see the light-shaded voxels. First digits of each ray signifies the index for the camera, and the second index is the index for that ray. Bottom: For each voxel we show how many different rays traverse that cell. The axes show the horizontal and vertical index of the voxels. Each voxel is identified by their horizontal and vertical indices.}
		\label{fig:intersection}
	\end{center} 
\end{figure}

In this article we will focus on the traversal of rays through a 3D array of voxels (in analogy to pixels, \textit{vo}lume \textit{el}ements) with constant spacing in each direction, but it can be generalized to voxels with varying widths, and even further to an octree where a (cubic) space is recursively subdivided into 8 sub-cubes in order to locally refine the 3D volume. For simplicity and didactic purposes, we will consider a rather simple scenario where there are only 3 cameras with a total of 7 rays, and where the situation can be visualized in 2D such that it is comprehensible, see fig.~\ref{fig:intersection}.

\begin{table*}[bth]
	\begin{center}
		\includegraphics[width=\textwidth]{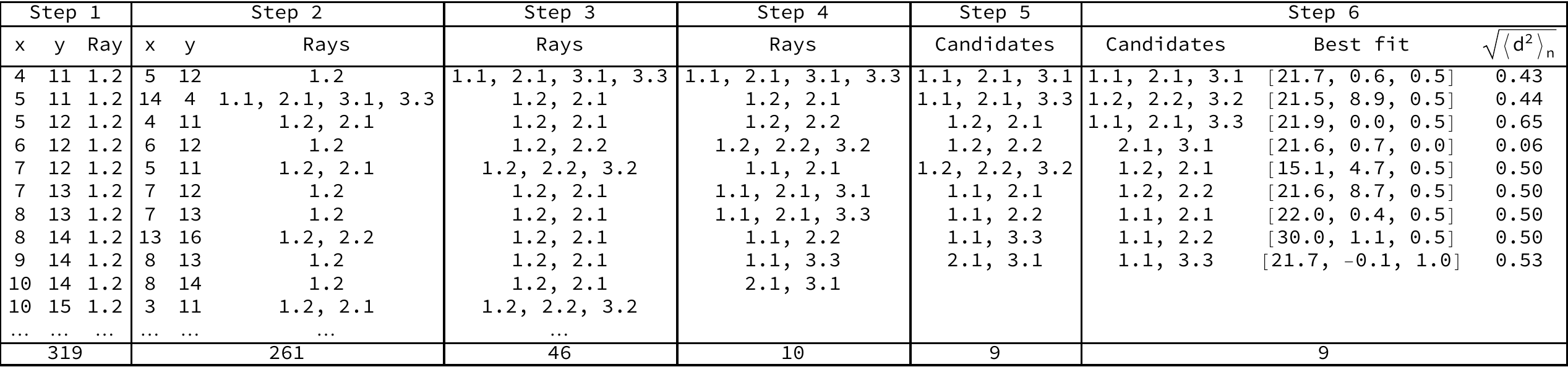}
		\caption{In step 1 we maintain a list of all the traversed voxels (denoted by the $x$ and $y$ indices) and the ray identifier in the \texttt{camera.ray} format. In step 2 the list of step 1 is grouped by the voxel indices $x$ and $y$, each voxel which is traversed by multiple rays will show multiple rays in the ray column. In step 3, we discard each voxel which is only traversed once because we need at least 2 rays (from different cameras) in order to get a match. In case there are more cameras we can require \textit{e.g.} at least rays from 3 different cameras. We also discard the voxel indices, as they are no longer needed. In step 4 we remove any duplicates from list the list of step 3. For each list of rays, we expand it in to all subsets 	\textit{i.e.} \texttt{1.1, 2.1, 3.1, 3.3} is expanded to two tuplets: \texttt{1.1, 2.1, 3.1} and \texttt{1.1, 2.1, 3.3}, after this `expansion' we remove again all the duplicates, the result is shown in the list of step 5. In step 6 we calculate the point which gives the position of the best match for each tuplet of rays, and the square root of the mean square distance is given (the matching error). This result is then sorted by number of rays (descending), and then by the error (ascending). Note that in real measurements the voxel indices are in 3D: $x$, $y$, and $z$. The last step is to pick matches from the top working down, while making sure each ray is only matched once.}
		\label{fig:grid}
	\end{center} 
\end{table*}

The first step of the algorithm is to traverse the rays through the voxels, this can be done very fast and is linear with the number of voxels in each direction, see \textit{e.g.}~Ref.~\cite{amanatides}. During this process we will maintain a list of all the voxels traversed, denoted by the two indices $x$ and $y$ (and $z$ in real experiments), along with the ray who traversed it, see step 1 of Table \ref{fig:grid}. Furthermore, we also add all the neighbours of the visited voxels (here we use a $\ell_1$ norm of 1 giving 6 neighbours: left, right, above, below, front, back). Such that these rays allow for some `play' during the matching. For our near-2D toy-problem shown in Fig.~\ref{fig:intersection}, we have 319 voxels that are visited. Our first element in the list is $x=4$, $y=11$, and $\text{Ray}=1.2$ meaning that ray 2 from camera 1 has visited the voxel with horizontal index 4, and vertical index 11. 

The second step is to gather, for each voxel, what rays have traversed through that voxel, see step 2 of Table~\ref{fig:grid}. One can see that a lot of voxels will have only a single ray that traversed them, but there are some, for example the second entry of the list, that is visited by four rays. This combining operation (akin to `group by' in SQL) can be done in $\mathcal{O}(n\log{}n)$ time where $n$ is the number of elements in the list; first the list is sorted by cell indices $x$ and $y$ in $\mathcal{O}(n\log{}n)$ time, and then one can walk through the sorted list `cutting' the list in `grouped' sublists in $\mathcal{O}(n)$ time. 

The third step is to remove entries from the list which are visited by only a single camera; for 3D matching we need information from at least 2 cameras in order to get a 3D coordinate. This step can be done in $\mathcal{O}(n)$ time. With our toy-problem we have now 46 elements in the list. Note that we can also discard the voxel indices as they are not needed any more, they were only needed in order for step 2 to combine the rays, or physically, to `compute' which rays are close to each other. The list of step 2 shows the cell indices only for explanation purposes but it could have been removed already during step 2. Note that the selection criterium can be generalized in the case one uses many cameras, one could \textit{e.g.}~require that at least 3 rays from different cameras are needed for a match, this would further prune the list. 

The fourth step is to remove any duplicates from the list of step 3. This can be done efficiently by first sorting the list of lists of rays in some canonical order (\text{e.g.}~small lists before long lists, and lists with equal length are ordered first by each first ray, then by the second ray, and so on, equivalent to a phone book where names have different lengths and the names are sorted first by the first character, then the second character and so on.), this sorting operation can be done in $\mathcal{O}(n\log{}n)$ time. Now we can again walk through the sorted list and only keep an element if the one before was not the same, this can be done in linear time $\mathcal{O}(n)$. Note that during step 2 we sorted the rays for each entry, this makes this sorting and deleting duplicates much easier. We are left with 10 entries in our list.

In the fifth step we extract all possible candidate from each of the set of rays. In general this means that the rays are grouped by camera, and then the Cartesian product is applied to get all the tuples of possible candidates. To exemplify, say that we have an entry: \texttt{1.1}, \texttt{1.2}, \texttt{2.1}, \texttt{3.1}, \texttt{3.2} this would give 2 possibilities for camera 1, 1 possibility for camera 2, and 2 possibilities for camera 3 or a total of 4 combinations of candidates ($\left\{\texttt{1.1},\texttt{2.1},\texttt{3.1}\right\}$, $\left\{\texttt{1.1},\texttt{2.1},\texttt{3.2}\right\}$, $\left\{\texttt{1.2},\texttt{2.1},\texttt{3.1}\right\}$, and $\left\{\texttt{1.2},\texttt{2.1},\texttt{3.2}\right\}$). For our toy-problem we can see that the first entry of step 4 is expanded in to 2 possible candidate matches (the first two entries of step 5). 

\begin{figure}[!bht]
	\begin{center}
		\includegraphics[width=0.8\columnwidth]{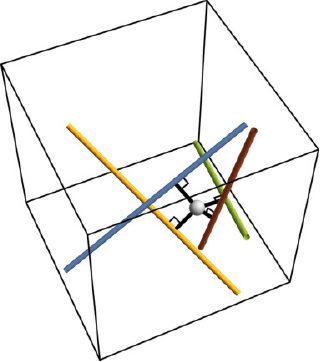}
		\caption{Example of closest point to \makered{4} rays (shown in red, green, blue\makered{, and yellow}) minimizing the sum of the square distances (shown in black) between the rays and the point (shown \makered{as a} gray \makered{sphere}).}
		\label{fig:bestfit}
	\end{center} 
\end{figure}

In the sixth and penultimate step we calculate the point for which the \makered{average} of the square distances is minimum for each candidate set of rays. This can be efficiently solved using a set of linear equations (3 of them to be precise, one for each coordinate) that can be solved using standard matrix algebra, see Fig.~\ref{fig:bestfit} for an illustration \makered{for the case of 4 rays}. For this point we calculate the root mean square value of the distances from this point to the rays (matching error); a smaller value means the rays cross more closely; a better and more probable match. Finally these candidates are sorted, first by number of rays in descending order (matches with more rays are more reliable than matches with less rays), and then by matching error in ascending order (smaller error is better). See the list in step 6 of Table \ref{fig:grid}.

The last step would be to walk through the list of candidates, starting from the top, and picking each of them for which the rays are not matched before. A na\"ive approach would be to remove all future entries which has one of the rays of the accepted candidate, however this would lead to $\mathcal{O}(n^2)$ time scaling which is unfavourable. We can do this more efficiently by keeping a hash map (hash table) which records the number of times a ray is used. This allows for fast insertion, looking up, and modification of the number of times a ray is used. This leads to an algorithm that scales with time as $\mathcal{O}(n)$ on average, and with a worst-case of $\mathcal{O}(n\log{}n)$. \makered{Note that the current implementation of the algorithm already allows for multiple matches per ray if needed; for a high particle density system particles can occlude the field of view such that particles appear to overlap in an image which then results in the creation of a single ray. Our code allows for matching each ray multiple time if needed. This, of course, should be used cautiously as it has the potential to result in so-called ghost particles. Different strategies can be implemented, prioritizing the number of cameras, the error, or the number of cases it is used. Also different strategies can be implemented such as the strategy proposed by Tan et al. \cite{tan2020} which includes the use of a so-called preference vector $P$. Additionally one could include the associated particle size in finding the best match in the selection procedure. The algorithm is flexible and can be optimized for different goals (best matches, most cameras used, least ghost \textit{et cetera}).}

As one can see, most of the steps have $\mathcal{O}(n\log{}n)$ time complexity or better, where $n$ is the number of traversed voxels. We can safely assume that the number of traversed voxels scales linearly with the number of rays (and independent of the number of cameras). We do, however, have to consider the size of the voxel. The voxels should be equal or larger than the maximum distance we allow as `error' for our matching. If we allow for a maximum distance of (say) \unit{1}{\milli \meter}, we need to make sure that all the voxels within a neighbourhood of \unit{1}{\milli \meter} from the ray are also traversed, this is done by also traversing the neighbours, see the light-shaded voxels in Fig.~\ref{fig:intersection}. It ensures us that matches with errors twice this \unit{1}{\milli \meter} will not occur, as two (or more) rays that are more than twice the distance away from each other will not traverse the same cells. It is therefore not a good idea to introduce a smaller voxel size than this distance in combination with a larger neigborhood of say 2 in $\ell_1$ distance in order to fulfil the requirement of maximum distance. This would just lead to more duplicate entries in the lists of steps 2 through 4.  For speed \makered{and memory} reasons we do have to make a compromise in the choice of our voxel spacing. A very fine spacing will give much better pruning in possible candidates (the list of step 5) but memory usage and time consumption will be high in the first steps as more cells are traversed; the $n$ in the aforementioned time complexity measures will increase inversely with the voxel size. However, if we make the voxel very large, say, in the extreme case, 1 voxel that fills the entire volume, then all the rays will traverse this single voxel, so the list of step 1, 2, 3, and 4 will be of length $\sum_i m_i$, however the length of step 5 will be $\prod_i m_i$, and will scale as $m^c$, where $m$ is the number of particles and $c$ the number of cameras; this is our na\"ive approach of trying out all combination of rays. And optimum is to be found with a voxel size (and maximum distance) for which the entries in steps 1 through 4 are manageable, while the number of candidates in step 5 is also kept manageable.

\section{Timing}
\begin{figure}[bht]
	\begin{center}
		\includegraphics{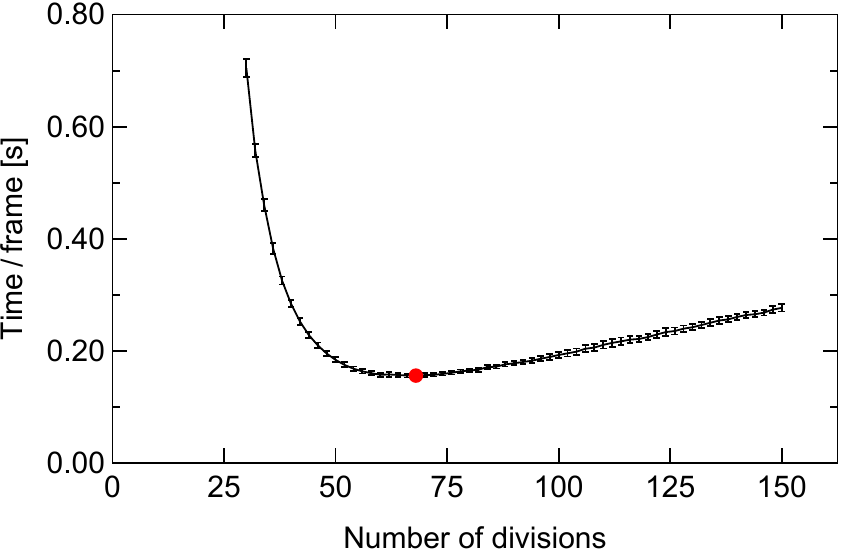}
		\caption{Time consumption as a function of the number of divisions of the voxel grid for 256 particles and 4 cameras. The fastest execution was found for \makered{68} divisions (shown in red), balancing the cost of traversing many voxels (large number of divisions) and having many candidates (small number of divisions). Error bars are based on repeated timings.}
		\label{fig:256timing4c}
	\end{center} 
\end{figure}

In order to find the optimum number of divisions, we will consider an artificial experiment with 256 randomly placed particles in a cubic volume with sides of length 1. We will add 4 virtual cameras, and position them around the volume in a \makered{tetrahedral} configuration. For now, we will consider a perfect arrangement of pinhole cameras such that the virtual rays are perfectly going through the virtual particles. For simplicity we will keep the voxels cubic, and vary the voxel size in all three direction simultaneously and time the duration of the execution, see Fig.~\ref{fig:256timing4c}. During step 3 we have only kept voxels which are visited by at least 3 cameras.

\begin{figure}[bht]
	\begin{center}
		\includegraphics{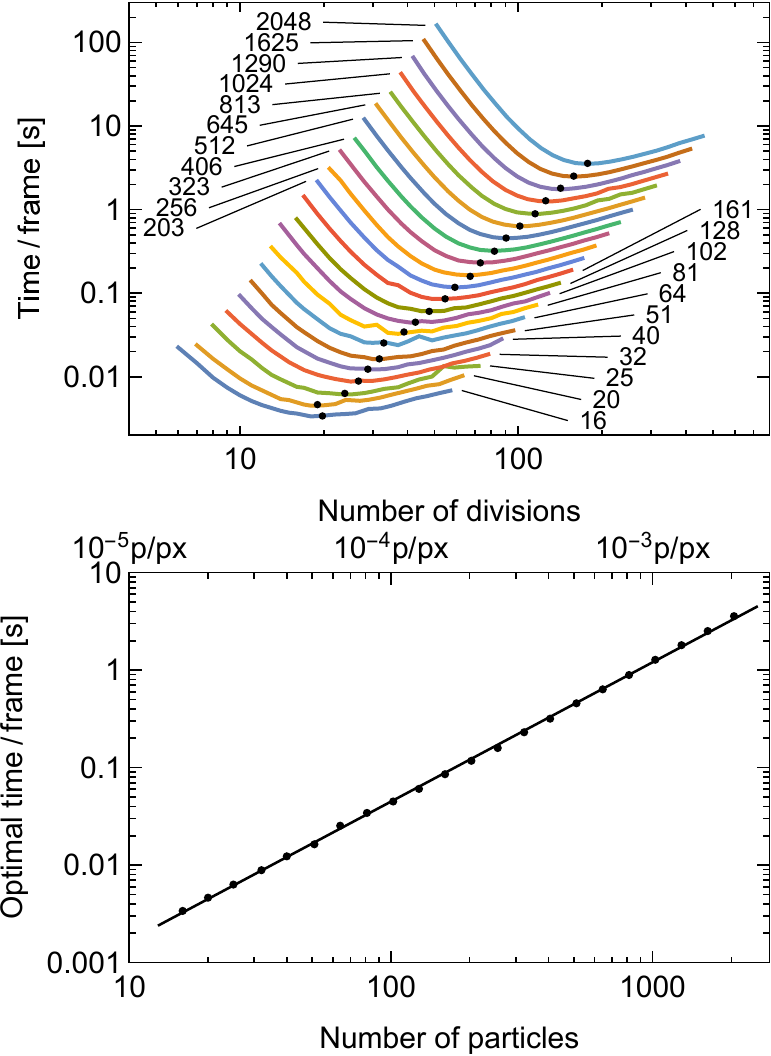}
		\caption{(Top) Timing as a function of number of divisions for several number of particles for 4 cameras. The number of particles are indicated in black for each dataset. For each dataset the minimum is marked with a black point. (Bottom) Optimal timing as a function of the number of particles for a 4 camera arrangement. The time is found to scale as \makered{$t \propto m^{1.429\pm0.006}$}. The prefactor will depend on \text{e.g.}~the \makered{computer}, the exact distribution of the particles, the implementation \textit{et cetera}. \makered{Top axes is the equivalent particle density for a typical 1 megapixel image, given in particles per pixel.}}
		\label{fig:divtimingparts}
	\end{center} 
\end{figure}

It can now clearly be observed that \makered{we find} a certain number of divisions of the volume (that determines the voxel size) for which the duration of the algorithm is minimum\makered{, in our configuration it is 68 divisions}. Such a minimum could be found automatically for a real measurement by e.g.~a golden-section search, before processing the entire recording. We can repeat this procedure for different number of particles, see Fig.~\ref{fig:divtimingparts}. One can see that the same balance can be found for all datasets of different number of particles. As expected, the optimal number of divisions ($d$) increases with the number of particles ($m$). We find that \makered{$d \propto m^{0.461\pm0.005}$} for a 95\% confidence level. The number of cells traversed, for optimal timing (and therefore divisions), scales therefore as \makered{$n \propto m^{1.461\pm0.005}$}.

In Fig.~\ref{fig:divtimingparts} the optimal computing time as a function of number of particles is shown. It is found that the time $t$ scales as \makered{$t \propto m^{1.429\pm0.006}$} for a 95\% confidence level. Which is close to the $m^{1.461}$ scaling that was predicted above. The scaling of $m^{1.429}$ means that doubling the number of particles results in only $2.7\times$ more processing power.

\section{Performance}
\begin{figure}[bht]
	\begin{center}
		\includegraphics{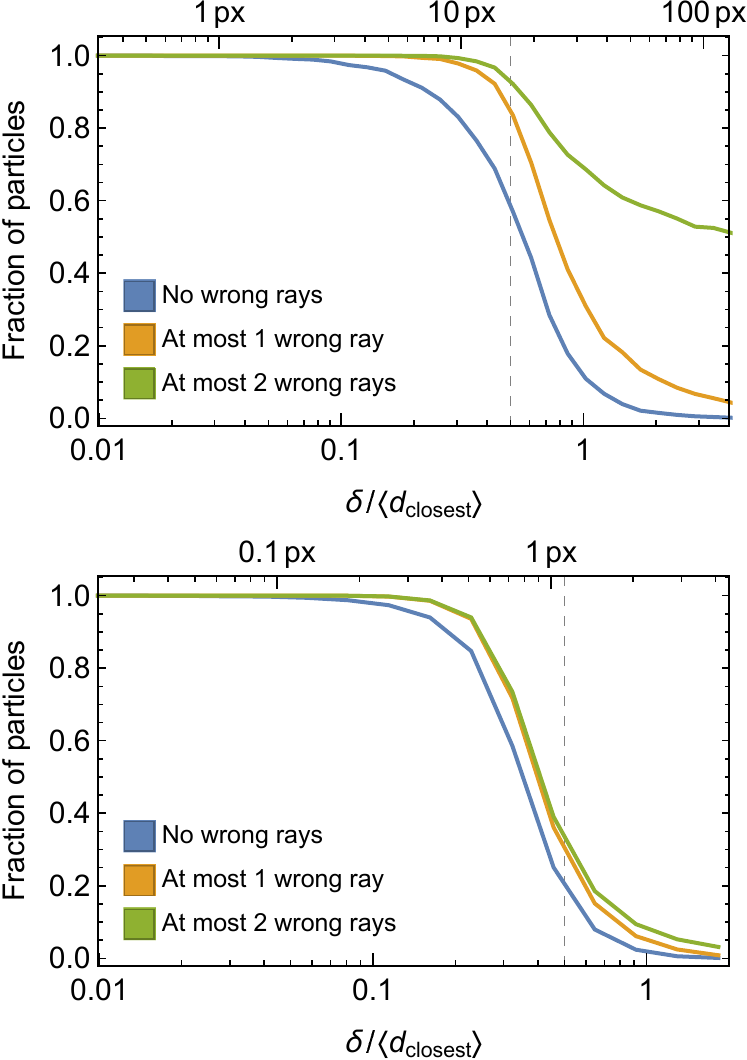}
		\caption{\makered{(Top)} Synthetic matching performance benchmark for 50 frames of 256 particles with random disturbances $\delta$ normalized by the average normal distance between adjacent particles $\left \langle d_{\text{closest}} \right \rangle$. 4 cameras are used in a tetrahedral configuration. $\delta / \left \langle d_{\text{closest}} \right \rangle = 0.5$ corresponds to the case that, on average, the random disturbance of particles are such that neighbouring point can touch. \makered{(Bottom) Same as above but for a single frame and 50000 particles. For both graphs the top axes are scaled such that the disturbances are expressed as the equivalent number of pixels for a typical 1 megapixel image for 256(Top) and 50000(Bottom) particles.}}
		\label{fig:disturbance}
	\end{center} 
\end{figure}

\subsection{\makered{General considerations}}
Next, the performance of the matching algorithm will be tested by randomly disturbing the particles in 3D for each camera. This causes the rays not to perfectly intersect, in order to mimic optical imperfections, which are unavoidable in a real PTV experiments. We artificially perturb the synthetic particles \makered{up to} a distance $\delta$ \makered{by generating a random vector inside a ball of radius $\delta$ (uniformly sampled in the volume of the ball). This is done} for each camera for the case of 256 particles situated inside a cubic region, observed by 4 cameras in a tetrahedral configuration. We compute the average distance between every particle and its closest neighbour as seen by the camera which is denoted by $\left \langle d_{\text{closest}} \right \rangle$. When $\delta/\left \langle d_{\text{closest}} \right \rangle \geq 0.5$ the particles can be disturbed so much that they can \makered{(on average)} start `touching'. In Fig.~\ref{fig:disturbance} we show the matching statistics for 50 frames of 256 random particles, perturbed for a variety of disturbances $\delta$. It can be seen that if the particles are mismatched 20\% of the mean inter-particle distance more than 90\% is still correctly matched. Note that this highly depends on the arrangement of the cameras, the shape of the measurement volume, and if the particles exhibit clustering. 

\begin{figure}[bht]
	\begin{center}
		\includegraphics{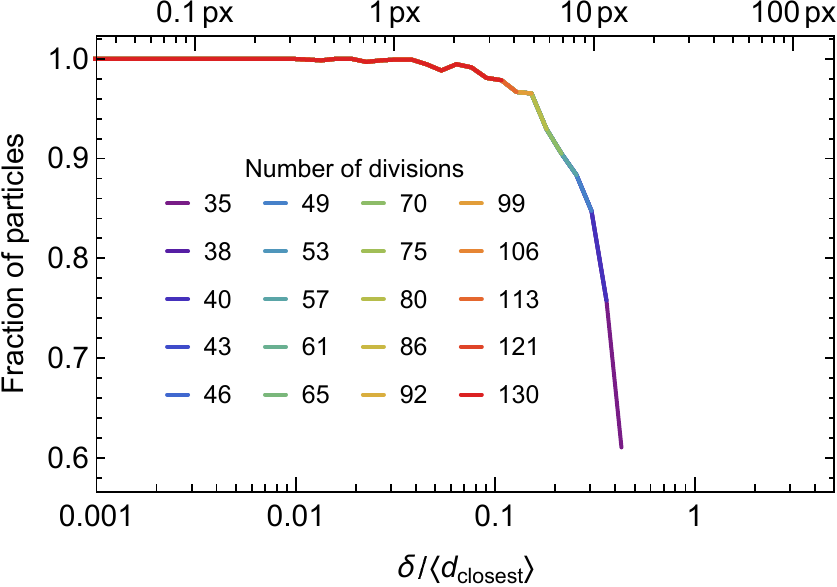}
		\caption{Accuracy of the matching as a function of the disturbance for a variety of divisions. For each disturbance $\delta$ and for each number of divisions we look at 10 frames of each 256 rays for each of the 4 cameras. The average performance of perfect matches is shown. All the curves perfectly overlap. Note that for large disturbances one can not choose too many divisions as the expanded voxels would not overlap in 3D (see Fig.~\ref{fig:intersection}). Top axis indicates the equivalent disturbance for a typical 1 megapixel image with 256 particles.}
		\label{fig:accuracydivs}
	\end{center} 
\end{figure}

\makered{To prove that our method is independent of the number of divisions we perform a synthetic test for a variety of disturbances $\delta$ for 10 frames of 4 cameras with 256 particles, for a variety of voxel sizes. As said before, the choice of voxel size has to be chosen carefully in the sense that it should not be smaller than the expected disturbance of the particles in an experiment. In such a case we can not guarantee that the traversed voxels overlap for each of the rays. So for a certain disturbance we have a lower limit on the voxel size (or an upper limit on the number of divisions). Note that for the case of perfect rays, without a disturbance, the number of divisions does not matter, the result is always the same, though the time and memory use can vary greatly. The results of our synthetic test that confirms our statement is shown in Fig.~\ref{fig:accuracydivs}. We see that all the curves overlap, such that the accuracy is independent of the number of the divisions. We note that each of the curves ends at a different $\delta$ such as to fulfill the requirement that the voxel size should not be smaller than the disturbance.}

\subsection{Comparison}\label{sec:comp}
\begin{figure}[bht]
	\begin{center}
		\includegraphics[]{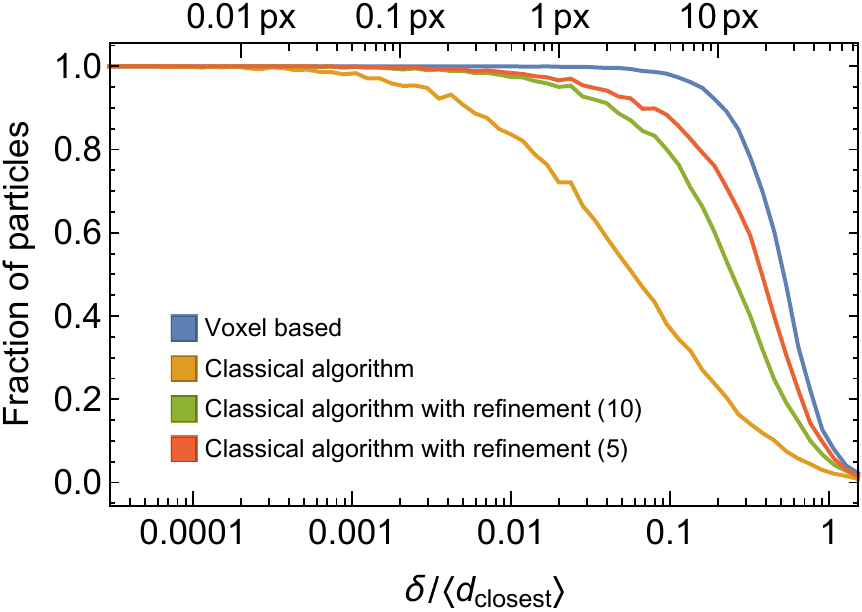}
		\caption{Synthetic matching performance for the case of 100 particles observed by 4 cameras in a tetrahedral configuration. Average performance of a 100 frames. Classical algorithm is described in Section \ref{sec:comp}. The refinements are done for the best 5 and the best 10 matches.}
		\label{fig:comparison}
	\end{center} 
\end{figure}

\makered{In this section we compare it to classic methods that rely on matching rays (or equivalently particles) based on pairs. Various different approaches and algorithms exist, but the general description is as follows: \circled{1} a ray (particle) is selected in one of the cameras, \circled{2} this ray is matched (either by projection on images or directly in 3D) to a ray or multiple rays on another camera, \circled{3} this is continued for each of the cameras until there is a reasonable match. This can be done by adding additional rays to each match so going from a single ray to pairs, to triplets, to quadruplets etc., or by looking for pairs of cameras and then combining these results. \circled{4} This match (and the corresponding rays (particles)) are excluded and the process is iterated until all rays are matched. This iterative nature of the algorithm has some downsides as the order of the rays which are chosen on the first cameras may influence the output, but also the order of the cameras can change the outcome. Several refinement strategies exist to (partially) negate this effect, but these can be computationally expensive. We compare our code to the above-described algorithm by selecting the `first' ray from camera 1, finding the best matching ray in camera 2, using this pair we select the best matching ray in camera 3 to form a triplet, and then to select the best ray in camera 4. These rays are then removed from the pool of rays, and the algorithm is repeated until no rays are left. We also implement an improvement on this algorithm, where, after this first run, only the best 10 (or 5) matches get selected, the remaining rays get shuffled, and then the algorithm is rerun to select another 10 (or 5) best matches, and this is repeated until no rays are left. Note that the performance of the algorithm depends on how the shuffling is done, and therefore the process is done for a 100 frames and the average performance is reported. We test both algorithms for 100 frames for the case of 4 cameras and 100 particles randomly positions in a spherical domain, see Fig.~\ref{fig:comparison}. It can be seen that the voxel-based matching algorithm gives best performance compared to the tested pair-based matching algorithms.}

\subsection{High particle count}
\makered{We also stress-test our method for the case of high number of particles like is done for related algorithms like \textit{shake-the-box} \cite{schanz2016shake,tan2020}. Note that in that method the tracking is done together with the matching, and it `works' directly on the images, while our method does not use information from previous frames, knowledge of the optics, or the camera arrangement and works solely on the rays. All our test were performed on a standard laptop which limits us to 50000 particles in the current implementation of the algorithm due to a peak memory consumption of \unit{29}{\giga\byte}. For 50000 particles randomly placed in a cube observed by 4 cameras in a tetrehedral configuration, we used $758$ divisions in each direction. The total number of voxels traversed is of the order of $850$ million, while the number of candidates matches is of the order 16 million. This also shows the difficulty of this method for high number of particles as all the traversed voxels have to be in memory. We show the performance of the method in the bottom of Fig.~\ref{fig:disturbance}. Note that for 50000 particles spread on a typical 1 megapixel image the average separation between particles is roughly \unit{2.2}{px}. Particles can be detected with sub-pixel accuracy, and the error can be smaller than \unit{0.1}{px} by utilizing the intensity of the neighbouring pixels to find an improved estimate of the centroid of the particle. Even for 0.4 pixel disturbances we find the performance to be above 90\%.}

\section{Memory}
\begin{figure}[bht]
	\begin{center}
		\includegraphics[width=\columnwidth]{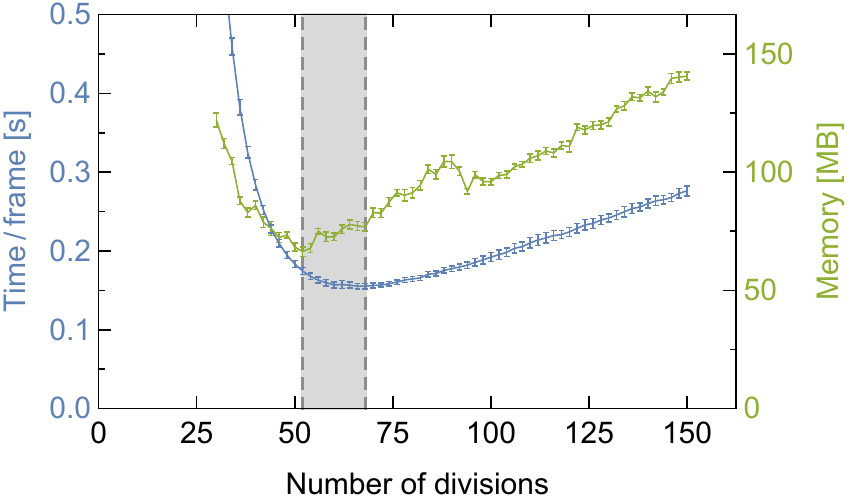}
		\caption{Memory and time usage for the case of 256 particles observed by 4 cameras for varying number of divisions. The dark gray region shows the optimal selection for the number of divisions. The user can either optimize for memory by selecting around \makered{52} divisions, or optimize for speed by selected around \makered{68} divisions in each direction. Going outside this region will increase both the time and memory needed to run the code. The error bars are based on repeated runs on the same hardware.}
		\label{fig:memory}
	\end{center} 
\end{figure}

\begin{figure*}[!bht]
	\begin{center}
		\includegraphics[width=\textwidth]{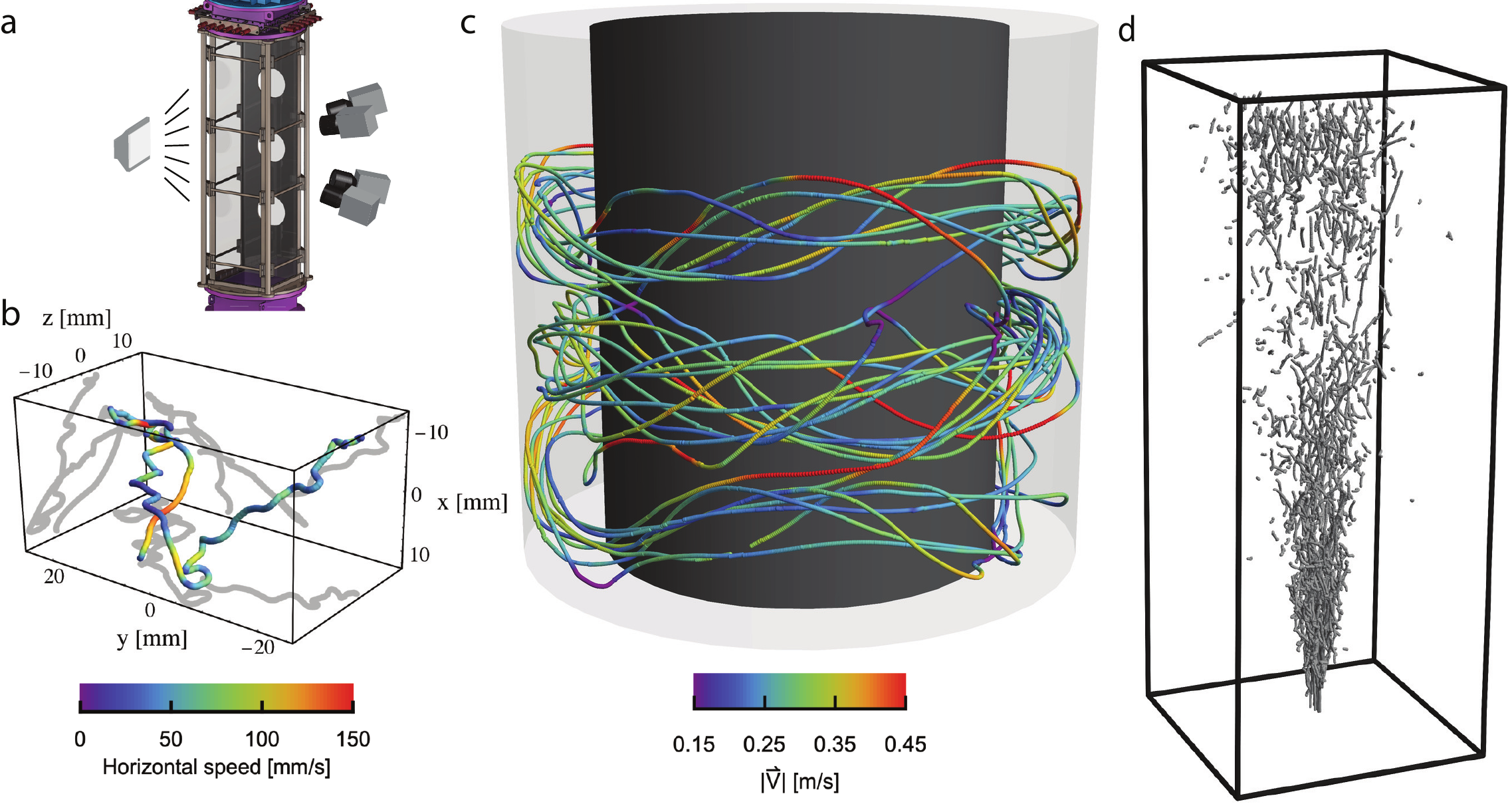}
		\caption{(a) Measurement section of the Twente Water Tunnel with a cross section of $\unit{45}{\centi\meter}\times\unit{45}{\centi\meter}$. Four Photron 1024PCI cameras view a common volume in the center of the tunnel. (b) Trajectory of a bubble (colored by its horizontal speed) tracked during 595 frames (corresponding to a duration of \unit{2.38}{\second}) at $\text{Re}_\lambda = 230$.  Figure adapted with permission from Mathai et al. \cite{mathai2018}. \d{Copyright 2018 by the American Physical Society.} (c) Trajectories of \unit{3}{\milli\meter} density-matched spheres (color by its speed $|\vec V |$) in counter-rotating Taylor Couette turbulence with $f_i=-0.4 f_o$ and $\text{Re} = (2\pi f_i r_i - 2\pi f_o r_o)(r_o-r_i)/\nu = 8\times10^4$. Data is obtained by combining 8 high-speed cameras. Inner cylinder has a diameter of \unit{150}{\milli\meter}. (d) Visualization of a 1000 tracks of a turbulent jet with particles obtained from tracer particles detected by 3 cameras. The box has sides of \unit{50}{\milli\meter}.} 
		\label{fig:examples}
	\end{center} 
\end{figure*}

We now focus on the memory usage. We set up a synthetic case of 256 particles seen by 4 cameras in a tetrahedral configuration. We vary the number of divisions over a wide range and monitor the time and memory used by the program, more specifically it is set-up to monitor the so-called maximum resident set size, see Fig.~\ref{fig:memory}. We observe that both the memory consumption and time consumption show an optimum. However, these are slightly shifted, which is expected as storing the traversed voxels and the candidates has a different memory ratio then the time-ratios for the time taken for traversing through the voxels and going through the candidate matches---the algorithms used have different memory and time dependencies. For the user it is best to select the number of divisions in the gray area, optimizing for memory (around \makered{52} divisions) or for speed (around \makered{68} divisions). These numbers will of course depend on the exact implementation and the \makered{programming} language(C++11), the compiler(GNU++11), the operating system(macOS Catalina), the hardware, the number of particles (256) and their arrangement, and the number of cameras(4) and their arrangement, where the values in parentheses are the parameters used for this publication.

\section{Examples}

The above matching algorithm has already been successfully applied to a variety of geometries with different number of cameras. We note that the tracking of the particles was done separately with standard particle tracking algorithms \cite{malik1993particle,ouellette2006quantitative}. We will go over several of the use-cases:

The code, as described above, has been used in the Twente Water Tunnel facility~\cite{poorte2002experiments}, which is a vertically oriented water tunnel that is \unit{8}{\meter} tall, which has an active grid to create near homogeneous isotropic stationary turbulence. It was used to track millimetric bubbles in 3D using 4 high-speed cameras, see Ref.~\cite{mathai2018} and Fig.~\ref{fig:examples}a for an overview of the measurement section and the camera arrangement. An example trajectory, colored by its horizontal speed, of one of the bubbles is shown in Fig.~\ref{fig:examples}b. 

The code was also slightly modified to work in the Taylor-Couette geometry. For this case the rays stop traversing once it hits the inner cylinder or leaves the setup again, and candidate matches are only allowed if they are located between the cylinders. Tracking particles in a Taylor--Couette geometry warrants full optical access from the side, we have therefore chosen to perform these experiments in the Boiling Twente Taylor--Couette facility \cite{huisman2015a} as the outer cylinder is constructed from transparent PMMA. To track particles around the inner cylinder a large number of cameras are needed, we have therefore used 8 high-speed cameras all around the inner cylinder. We have selected 30 large trajectories for the case of counter-rotating cylinders $f_i=-0.4f_o$ and $\text{Re}=8\times10^4$ with finite-size neutrally-buoyant particles of $\unit{3}{\milli\meter}$ diameter to be visualized, see Fig.~\ref{fig:examples}c. The color of the trajectories represent the velocity magnitude of the particles.

Another use-case is for the measurements of a turbulent jet. Here 3 high-speed cameras were used to track neutrally-buoyant tracer particles of $\unit{250}{\micro\meter}$ in size, see Fig.~\ref{fig:examples}d. Lastly, the algorithm is also already used in Rayleigh-B\'enard convection experiments to track tracer particles and in the Lagrangian Exploration Module \cite{LEM} at the \'Ecole normale sup\'erieure in Lyon to track neutrally-buoyant particles for which publications are in progress.

\section{Conclusion}
In conclusion, we have explained a new \makered{fully scalable (with regard to the number of cameras used)} algorithm for finding matches of rays, the results of which can be used for Particle Tracking Velocimetry (PTV) or 3D reconstruction. The main advantage over other algorithms is that the order of cameras and the order of the rays (detected particles in 2D) do not influence the outcome. Another advantage is that it will first find the globally best match, rather than a greedy algorithm (locally optimal choice). Moreover, the algorithms shows good \makered{time} scaling as the number of cameras and particles increase. We have shown several use-cases with different geometries and with different number of high-speed cameras, however this algorithm is not limited to those geometries, as it could also be used to track particles (bubbles, droplets) in geometries such as pipes, von K\'arm\'an flow, oscillating grid turbulence setups, or the V-ONSET \cite{vonset}.

\newpage

\begin{acknowledgments}
This work was financially supported by the Zwaartekracht programme Multiscale Catalytic Energy Conversion (MCEC), which is part of The Netherlands Organisation for Scientific Research (NWO) and by the Project IDEXLYON of the University of Lyon in the framework of the French program  ``Programme Investissements d'Avenir'' (ANR-16-IDEX-0005). We thank Thomas Basset, David Dumont, Luuk Blaauw, Varghese Mathai, Chao Sun, Bianca Viggiano, and Romain Volk for feedback on, discussion of, and testing of the method and providing their experimental data for the example visualisations.
\end{acknowledgments}

\section{Data availability Statement}
The data that support the findings of this study are available from the corresponding author upon reasonable request.


\begin{thebibliography}{10}

\bibitem{maas}
H.~G. Maas, A. Gruen and D. Papantoniou, Experiments in Fluids {\bf 15},  133
  (1993).

\bibitem{virant}
M. Virant and T. Dracos, Measurement science and technology {\bf 8},  1539
  (1997).

\bibitem{ott}
S. Ott and J. Mann, Journal of Fluid Mechanics {\bf 422},  207  (2000).

\bibitem{bourgoin2006}
M. Bourgoin, N.~T. Ouellette, H. Xu, J. Berg and E. Bodenschatz, Science {\bf
  311},  835  (2006).

\bibitem{hartley2003}
R. Hartley and A. Zisserman, {\em Multiple view geometry in computer vision}
  (Cambridge university press, 2003).

\bibitem{zhang1995}
Z. Zhang, R. Deriche, O.~D. Faugeras and Q. Luong, Artificial Intelligence {\bf
  78},  87  (1995).

\bibitem{hartley1995}
R.~I. Hartley,  in {\em Proceedings of IEEE International Conference on
  Computer Vision}, IEEE (PUBLISHER, 1995), pp.\ 882--887.

\bibitem{criminisi2001}
A. Criminisi, {\em Accurate Visual Metrology from Single and Multiple
  Uncalibrated Images.}, {\em Distinguished Dissertation Series}
  (Springer-Verlag London Ltd., 2001).

\bibitem{basanta2013}
J. Basanta, O. Kazuo and N. Kazuo, International Journal of Innovative
  Computing, Information and Control {\bf 9},  5  (2013).

\bibitem{machicoane}
N. Machicoane, A. Aliseda, R. Volk and M. Bourgoin, Review of Scientific
  Instruments {\bf 90},  035112  (2019).

\bibitem{tsai}
R. Tsai, IEEE Journal on Robotics and Automation {\bf 3},  323  (1987).

\bibitem{mathai2018}
V. Mathai, S.~G. Huisman, C. Sun, D. Lohse and M. Bourgoin, Phys. Rev. Lett.
  {\bf 121},  054501  (2018).

\bibitem{bib:risoeReport}
J. Mann, S. Ott and J.~S. Andersen, Technical Report No.~Riso-R-1036 (EN),
  Risoe National Laboratory, Roskilde, Denmark (unpublished).

\bibitem{amanatides}
J. Amanatides, A. Woo {\it et~al.},  in {\em Eurographics} (-, 1987), No.~3,
  pp.\ 3--10.

\bibitem{tan2020}
S. Tan, A. Salibindla, A.~U.~M. Masuk and R. Ni, Experiments in Fluids {\bf
  61},  47  (2020).

\bibitem{schanz2016shake}
D. Schanz, S. Gesemann and A. Schr{\"o}der, Experiments in fluids {\bf 57},  70
   (2016).

\bibitem{malik1993particle}
N. Malik, T. Dracos and D. Papantoniou, Experiments in fluids {\bf 15},  279
  (1993).

\bibitem{ouellette2006quantitative}
N.~T. Ouellette, H. Xu and E. Bodenschatz, Experiments in Fluids {\bf 40},  301
   (2006).

\bibitem{poorte2002experiments}
R. Poorte and A. Biesheuvel, J. Fluid Mech. {\bf 461},  127  (2002).

\bibitem{huisman2015a}
S.~G. Huisman, R.~C.~A. van~der Veen, G.-W.~H. Bruggert, D. Lohse and C. Sun,
  Review of Scientific Instruments {\bf 86},    (2015).

\bibitem{LEM}
R. Zimmermann, H. Xu, Y. Gasteuil, M. Bourgoin, R. Volk, J.-F. Pinton and E.
  Bodenschatz, Review of Scientific Instruments {\bf 81},  055112  (2010).

\bibitem{vonset}
A.~U.~M. Masuk, A. Salibindla, S. Tan and R. Ni, Review of Scientific
  Instruments {\bf 90},  085105  (2019).

\end{thebibliography}

\end{document}